\documentclass[conference]{IEEEtran}
\IEEEoverridecommandlockouts

\usepackage{cite}
\usepackage{amsmath,amssymb,amsfonts}
\usepackage{algorithmic}
\usepackage{graphicx}
\usepackage{textcomp}
\usepackage{xcolor}
\usepackage{soul}
\usepackage{hyperref}
\usepackage{comment}
\usepackage{siunitx}
\def\BibTeX{{\rm B\kern-.05em{\sc i\kern-.025em b}\kern-.08em
    T\kern-.1667em\lower.7ex\hbox{E}\kern-.125emX}}
\begin{document}

\title{Graph Neural Network-Based Predictor for Optimal Quantum Hardware Selection \vspace{-10pt}
}

\author{\IEEEauthorblockN{Antonio Tudisco\IEEEauthorrefmark{1},  Deborah Volpe\IEEEauthorrefmark{2}, Giacomo Orlandi\IEEEauthorrefmark{1}, and Giovanna Turvani\IEEEauthorrefmark{1} }
\IEEEauthorblockA{\IEEEauthorrefmark{1}Department of Electronics and Telecommunications,
Politecnico di Torino
Italy\\
\IEEEauthorrefmark{2}Istituto Nazionale di Geofisica e Vulcanologia, Rome, Italy.
\\
 \href{mailto:antonio.tudisco@polito.it}{antonio.tudisco@polito.it},
   \href{mailto:deborah.volpe@ingv.it}{deborah.volpe@ingv.it}
  \href{mailto:giacomo.orlandi@polito.it}{giacomo.orlandi@polito.it},
\href{mailto:giovanna.turvani@polito.it} {giovanna.turvani@polito.it}}\vspace{-30pt}} 

\maketitle

\begin{abstract}
The growing variety of quantum hardware technologies, each with unique peculiarities such as connectivity and native gate sets, creates challenges when selecting the best platform for executing a specific quantum circuit. This selection process usually involves a brute-force approach: compiling the circuit on various devices and evaluating performance based on factors such as circuit depth and gate fidelity. However, this method is computationally expensive and does not scale well as the number of available quantum processors increases.

In this work, we propose a Graph Neural Network (GNN)-based predictor that automates hardware selection by analyzing the Directed Acyclic Graph (DAG) representation of a quantum circuit. Our study evaluates 498 quantum circuits (up to 27 qubits) from the MQT Bench dataset, compiled using Qiskit on four devices: three superconducting quantum processors (IBM-Kyiv, IBM-Brisbane, IBM-Sherbrooke) and one trapped-ion processor (IONQ-Forte). Performance is estimated using a metric that integrates circuit depth and gate fidelity, resulting in a dataset where 93 circuits are optimally compiled on the trapped-ion device, while the remaining circuits prefer superconducting platforms.
By exploiting graph-based machine learning, our approach avoids extracting the circuit features for the model evaluation but directly embeds it as a graph, significantly accelerating the optimal target decision-making process and maintaining all the information. Experimental results prove 94.4\% accuracy and an 85.5\% F1 score for the minority class, effectively predicting the best compilation target.\\
The developed code is  publicly available on GitHub (\url{https://github.com/antotu/GNN-Model-Quantum-Predictor}).
\end{abstract}


\vspace{-3pt}
\section{Introduction}
Quantum computing has arisen as a promising paradigm for solving computationally intractable problems, with applications spanning optimization, cryptography, and machine learning \cite{Preskill2018,Biamonte2017,Montanaro2016}. However, the variety of quantum hardware architectures --- each characterized by distinct qubit connectivity, native gate sets, and operational constraints~\cite{Murali2019,tannu2019not} --- presents a significant challenge in selecting the \textbf{optimal platform for executing a given quantum circuit}. For example, superconducting qubits offer fast gate execution but are constrained by limited connectivity, whereas trapped-ion devices can provide higher qubit connectivity but suffer from slower gate operations~\cite{Wright2019,Debnath2016}. Moreover, hardware variations in coherence times, gate fidelities, and noise models further complicate the selection process.

Quantum circuit execution on different platforms requires exhaustive evaluation since a circuit is compiled and tested on multiple devices to evaluate performance metrics such as circuit depth and gate fidelity. This brute-force approach, while effective, is \textbf{computationally expensive} and does not scale well as the number of available quantum processors increases \cite{quetschlich2024mqtpredictor}. As quantum computing ecosystems expand, a more efficient and \textbf{automated methodology} is needed to assist users --- especially those without deep expertise in quantum hardware --- in selecting the most suitable quantum processor for their applications.

In this work, we propose a \textbf{Graph Neural Network} (\textbf{GNN})-based approach to predict the optimal quantum hardware for a given quantum circuit. Our method exploits the \textbf{Directed Acyclic Graph} (\textbf{DAG}) representation of circuits, embedding them directly into a graph-based model without requiring manual feature extraction. By training on a dataset of \textbf{498 circuits} from the MQT Bench \cite{quetschlich2023mqt}, compiled on three superconducting devices (IBM-Kyiv, IBM-Brisbane, IBM-Sherbrooke) and one trapped-ion processor (IONQ-Forte), we prove that our approach effectively learns structural relationships within circuits, enabling accurate and scalable hardware selection.

Experimental results show that our model achieves 94.4\% accuracy in predicting the best execution platform in terms of outcomes quality, with an F1 score of 85.5\% for the minority class. By \textbf{automating the quantum hardware selection process}, our framework avoids the need for exhaustive compilation evaluations, significantly accelerating decision-making while maintaining high prediction accuracy. This work integrates machine learning into quantum computing workflows, making quantum resources more accessible and efficiently utilized.

The rest of the article is organized as follows.  In Section~\ref{sec:TheoreticalFoundation}, theoretical foundations behind the work are discussed.  Section~\ref{sec:RelatedWorks} reviews existing approaches leveraging machine learning in quantum compilation and hardware selection. Section~\ref{sec:Implementation} describes our methodology, including dataset construction and the GNN-based model. Section~\ref{sec:Results} presents experimental evaluations, and Section~\ref{sec:Conclusions} concludes the paper with future research directions.
\begin{figure*}[ht]
    \centering    \includegraphics[width=1\linewidth]{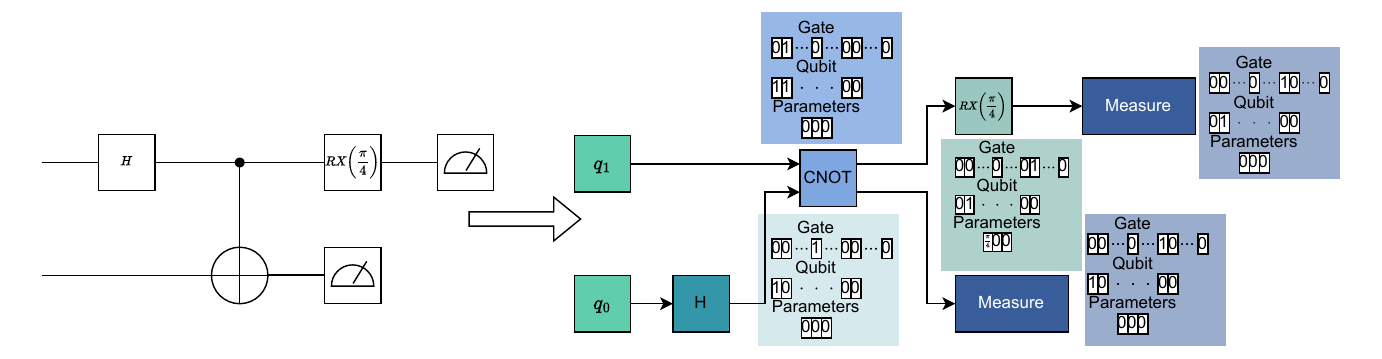} \vspace{-10pt}
    \caption{Representation on how to map a quantum circuit as a DAG \vspace{-150pt}}
    \label{fig:DAG}
\end{figure*}

\vspace{-3pt}
\section{Theoretical Foundation} \label{sec:TheoreticalFoundation}
This section briefly presents the \textbf{DAG} representation of a quantum circuit and, Machine Learning (ML) basis and the characteristic of \textbf{GNNs}.

\subsection{Direct Acyclic Graph}
Any quantum circuit can be represented as DAG, where nodes without incoming edges represent the circuit’s inputs --- i.e. the qubits ---, and each subsequent node corresponds to a quantum operation --- i.e. the quantum gates. Each operation node is characterized by a feature vector encoding details about the type of operation/gate, the involved qubits, and the associated parameters such as angles in rotational gates. This DAG-based representation is commonly employed during the \textbf{quantum circuit compilation process}, for example in the Qiskit transpiler \cite{meijer2025comparison}.
The key advantage of this representation is that it compactly \textbf{captures all the characteristics of the quantum circuit} to be executed.

Figure~\ref{fig:DAG} presents an example of a quantum circuit represented in DAG format. The qubits correspond to the nodes without incoming edges, while the remaining nodes represent operations. Each of them is characterized by a vector of three features: the gate type (represented using one-hot encoding), the qubits involved, and any associated parameter. The parameter field is empty for all operations except the rotational gate  $R_x$, for which it contains the rotation angle.
\begin{figure}[ht]
    \centering    \includegraphics[width=1\linewidth]{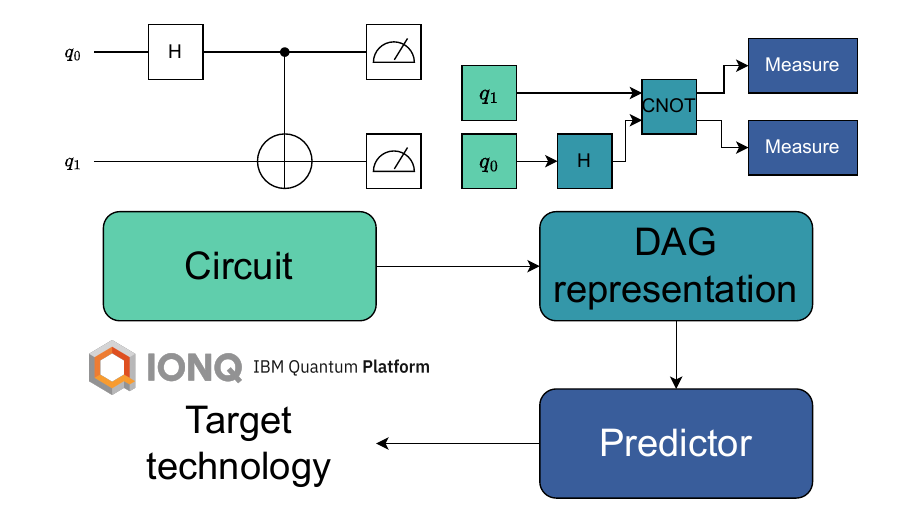}\label{-40pt}
    \caption{Workflow for the prediction of the best technology to exploit for quantum circuit execution. \label{-1000pt} }
    \label{fig:workflow}
\end{figure}

\subsection{Graph Neural Network}
GNNs \cite{liang2022survey} represent a class of deep learning architectures suitable for graph-structured data. These models transform the intrinsic features of nodes, edges, and entire graphs by systematically aggregating and propagating information along the edges, preserving symmetries such as permutation invariance.

Among the several models within this framework, \textbf{Graph Convolutional Networks} (\textbf{GCNs}) \cite{kipf2017semisupervisedclassificationgraphconvolutional} and \textbf{Graph Attention Networks} (\textbf{GATs}) \cite{veličković2018graphattentionnetworks} are two of the most common models.

GCNs extend the concept of convolution from grid-structured data, such as images, to graphs. In this case, the convolution operation is executed by aggregating feature information from a node’s local neighborhood. 

On the other hand, GATs include an attention mechanism into the neighborhood aggregation process. Rather than assigning equal importance to each neighboring node, GATs compute dynamic attention coefficients, emphasizing more relevant neighboring features during the aggregation phase.

Recent advancements have explored hybrid architectures that integrate the stable local feature aggregation of GCNs with the adaptive, context-sensitive weighting mechanisms inherent in GATs. Such integrated models aim to exploit the complementary strengths of both approaches, achieving enhanced performance.

The most important hyper-parameters for the GCN and the GAT are the number of input channels and, for the GAT, the number of attention heads.

\section{Related works} \label{sec:RelatedWorks}
ML is increasingly being employed to \textbf{automate processes}, especially in the field of Electronic Design Automation (EDA), where it has become very popular to automatize processes that spanning from formal verification to logic optimization, placing, and routing. 
Significant examples are \cite{10.1145/3224206}, which proposes a supervised machine learning approach for formal verification that predicts the optimal formal engine for a given design property, \cite{haaswijk2018deep}, which introduces a deep reinforcement learning method for automatic logic optimization in digital circuits, \cite{10213402}, that utilizes GNNs to estimate post-implementation Quality of Results from the pre-schedule control data flow graph representation of an High Level Synthesis (HLS) design targeting Field-Programmable Gate Arrays (FPGA) implementations, \cite{10.1145/3400302.3415690}, that presents a deep reinforcement learning framework designed to optimize the placement parameters in a commercial EDA tool, and \cite{10.1145/3508352.3549346}, that proposes an approach that exploit the GNNs to improve the prediction of routing congestion.

Based on these successes in EDA, recent research has started exploring the use of machine learning techniques for automating tasks within the quantum computing stack. In particular, several approaches have investigated automated compilation optimization. Among them, the \textbf{MQT Predictor} framework, presented in \cite{quetschlich2023predicting,quetschlich2024mqtpredictor}, introduces a supervised machine learning approach to predict the most suitable quantum hardware for a given circuit, alongside a reinforcement learning-based strategy for optimizing compiler parameters. Their method addresses device selection and compilation by evaluating a wide range of hardware and software configurations, aiming to reduce manual tuning and improve execution efficiency across heterogeneous quantum platforms. While this study focuses on classification-based predictions exploiting extensive benchmarking, our work introduces a fundamentally different approach based on graph-based machine learning. Instead of treating quantum circuit selection as a flat, feature-based classification task, \textbf{we represent circuits as DAGs and employ GNNs to capture the inherent structural relationships among quantum gates and qubits}. This graph-based representation allows our model to learn topology-aware embeddings that reflect the full circuit architecture without discarding structural details through manual feature extraction. \textbf{Unlike previous approaches that lean on a limited set of engineered features, our method retains all relevant information from the original quantum circuit}, enabling more expressive modeling and improved generalization across diverse circuit types. Furthermore, while earlier work focuses primarily on tuning compilation strategies or hardware-software co-optimization, our approach targets the circuit-to-hardware matching problem, ensuring that circuits are mapped to the most suitable device based on structural compatibility. Using GNNs over circuit DAGs improves scalability and adaptability as quantum hardware architectures evolve.

\begin{figure}
    \centering
\includegraphics[width=\linewidth]{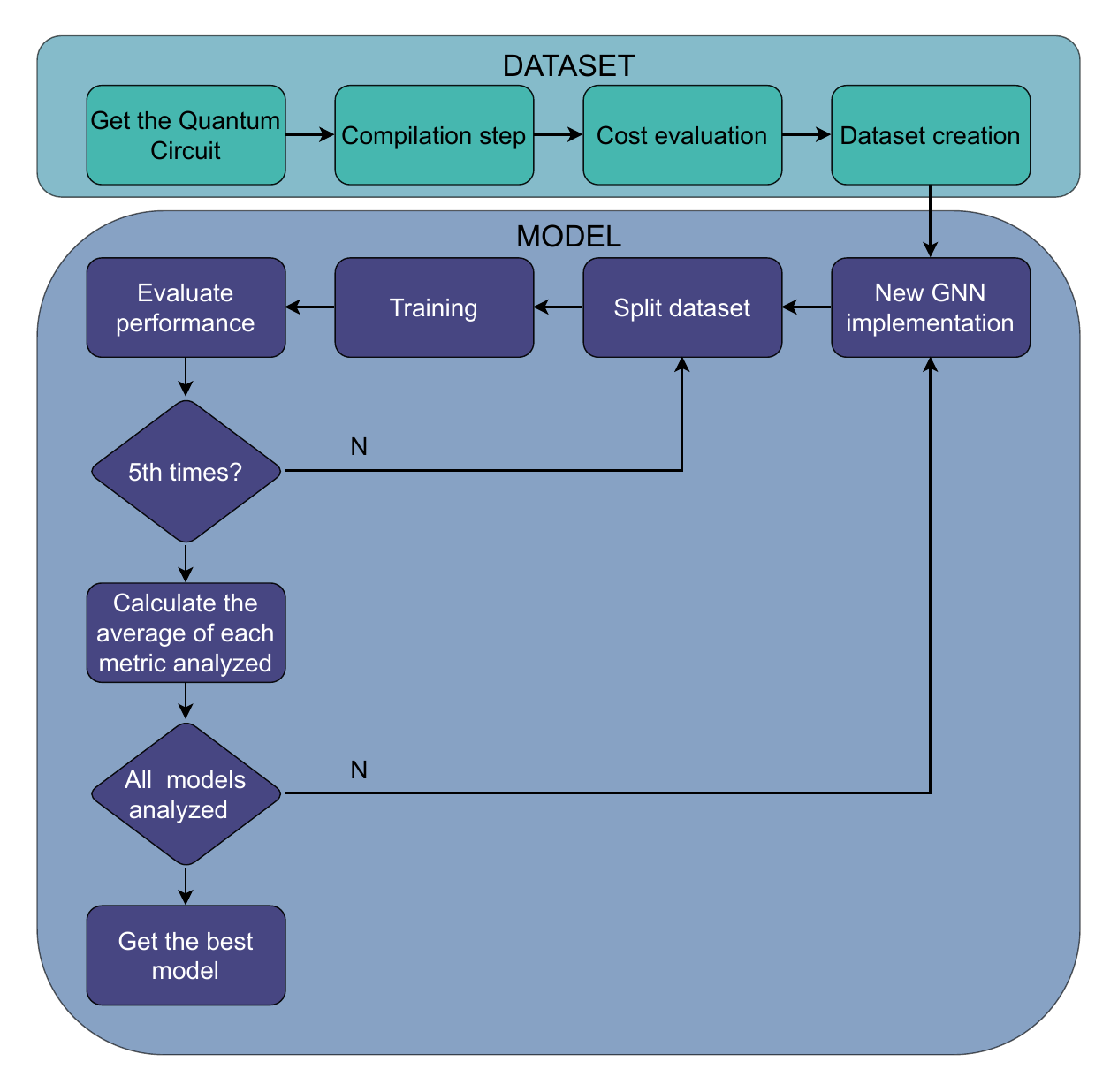}  \vspace{-20pt}
    \caption{Methodology followed to implement our predictor. \vspace{-150pt}}
    \label{fig:Step-Predictor}
\end{figure}

\section{Implementation}\label{sec:Implementation} 

This work aims to automate the process of selecting the optimal quantum device on which to run a quantum circuit. To this end, we first defined our dataset, which involves pre-processing and labeling operations and a prior analysis to evaluate the limitations of manual feature extraction, and then we trained several GNNs on it.
The methodology employed to implement our predictor is represented in Figure \ref{fig:Step-Predictor}, where the two main blocks, the dataset and model definitions, are highlighted.
\subsection{Dataset}

The dataset was constructed using 498 quantum circuits up to 27 qubits from the MQT Bench \cite{quetschlich2023mqt} tool. These circuits were compiled using the Qiskit compiler varying the optimization level from 0 to 3 and targeting three superconducting IBM devices (IBM-Kyiv, IBM-Brisbane, and IBM-Sherbrooke) and a trapped-ion device (IONQ-Forte).

The three superconducting devices are characterized by 127-qubit, and the processor employed is Eagle. Its basis gates include ECR, ID, RZ, SX, and X. They also have an average $T_1$ of 260-270 \si{\micro \second} and an average $T_2$ of approximately 150 \si{\micro \second}.

The IONQ-Forte is a 36-qubit trapped ion device, characterized by an all-to-all connectivity and a high fidelity (99.6\% for the 2 qubit quantum gates and 99.98\% for the single qubit gate) and long coherence times ($T_1$ in the range 10-100 \si{\second}, $T_2$ about 1 \si{\second}).

From the compiled circuits on each device, we extract the fidelity of each gate and the resulting circuit depth. These values are used for evaluating the device effectiveness for the execution of the circuit, through the following cost function \cite{https://doi.org/10.1002/qute.202300128}:
\begin{equation}
- D \log K - \sum_i \log F_i\, ,
\end{equation}
where $D$ is the depth of the compiled circuit, $K$ is a constant equal to the average of the highest and lowest gate fidelities in the compiled circuit on the given device, and $F_i$ is the fidelity of the $i^{\text{th}}$ gate. This metric balances circuit depth and overall fidelity --- for which, considering the high fidelity of the quantum gates, the trapped-ion technology is particularly advantageous.

The original benchmark circuits considered were then described in the  DAG form, as in the example of Figure \ref{fig:DAG}. In this case, the feature vector of each node is composed of 66 binary elements: the first 36 elements use one-hot encoding to specify the gate type, the next 27 identify the target qubit, assuming circuits of up to 27 qubits, and  final 3 elements represent the gate’s angle parameters. Each DAG is labeled with the optimal technology for running the corresponding quantum circuit, considering the device and compilation settings that yield the lowest overall cost.

Finally, the problem was reformulated as a binary classification task to determine the optimal technology for executing each circuit. Circuits where the IONQ-Forte (trapped-ion technology) is the best were assigned to class 0, while those for which superconducting technology is preferable were assigned to class 1. Consequently, the dataset comprises 93 instances in class 0 (trapped-ion) and 405 instances in class 1 (superconducting). 

Treating the quantum circuit in graph form and considering the GNN models avoid the explicit feature extraction step, employed in  \cite{quetschlich2024mqtpredictor}. This strategy is expected to significantly improve the classification performance since a manual features extraction can lack of relevant information for the classification pattern identification. 

\begin{figure}[ht]
    \centering
    \includegraphics[width=\linewidth]{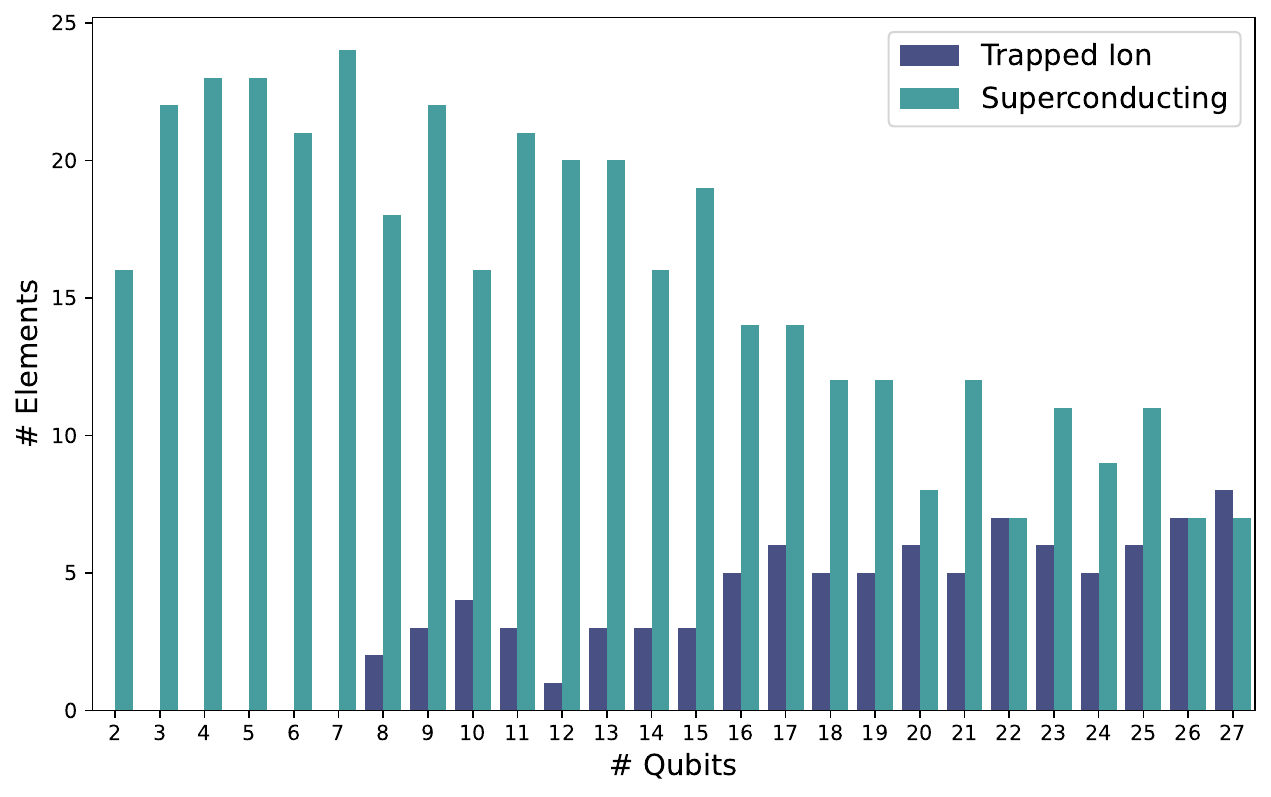}
    \caption{Distribution between the two classes of the input circuits based on the number of qubits.}
    \label{fig:DistibutionVSQubits}
\end{figure}

\subsubsection{Initial Analysis}

First, we analyzed how the circuit characteristics such as depth, gate number, or number of qubits affect the performance on the considered device and so the technology choice. Then, the distribution of the target device classes across the various circuit was examined. Specifically, the correlation between the number of qubits in the quantum circuit and the classes distribution was observed.

As shown in Figure \ref{fig:DistibutionVSQubits}, for circuits with less than 8 qubits, the optimal device is typically the superconducting one. This suggests that, in smaller circuits, the higher gate fidelity outweighs the disadvantage of greater circuit depth. However, as the number of qubits increases, this trade-off shifts, making trapped-ion technology a reasonable alternative considering the metric chosen.

\begin{figure}[ht]
    \centering
    \includegraphics[width=\linewidth]{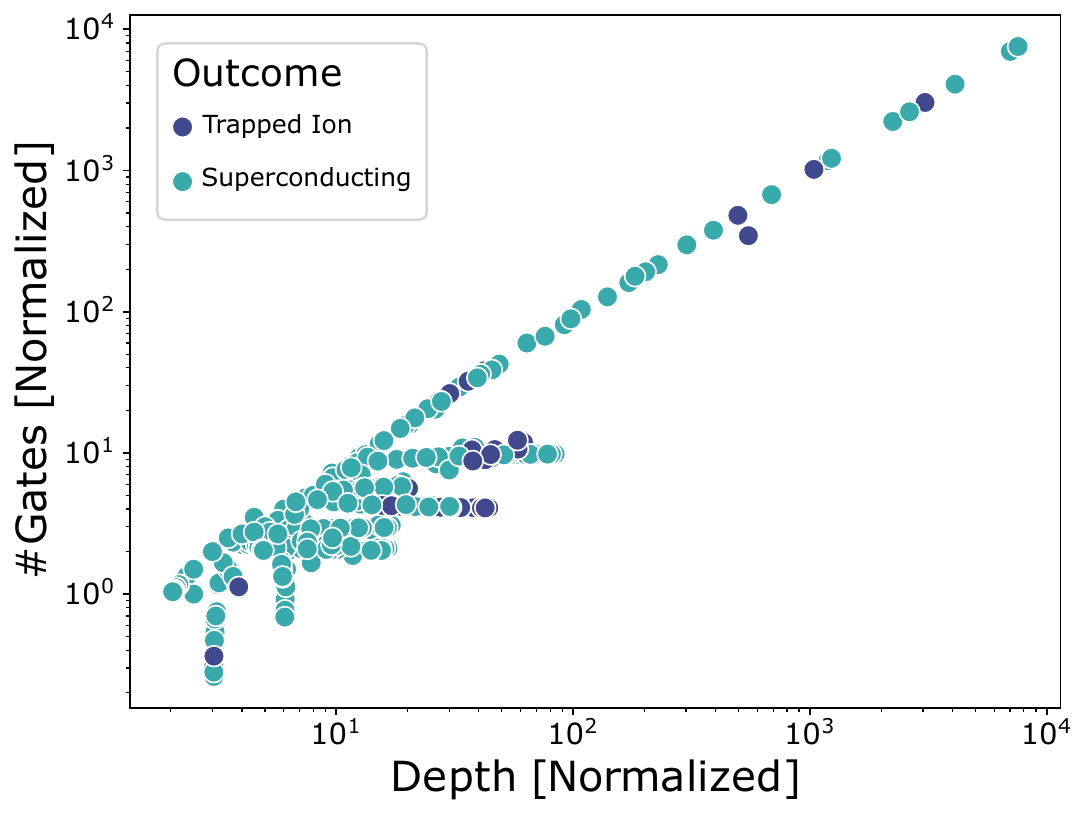}
    \caption{Distribution of circuits based on circuit depth and number of nodes, normalized by the number of qubits.}
    \label{fig:depth_nodes_normalized}
\end{figure}

\begin{figure}[ht]
    \centering
    \includegraphics[width=\linewidth]{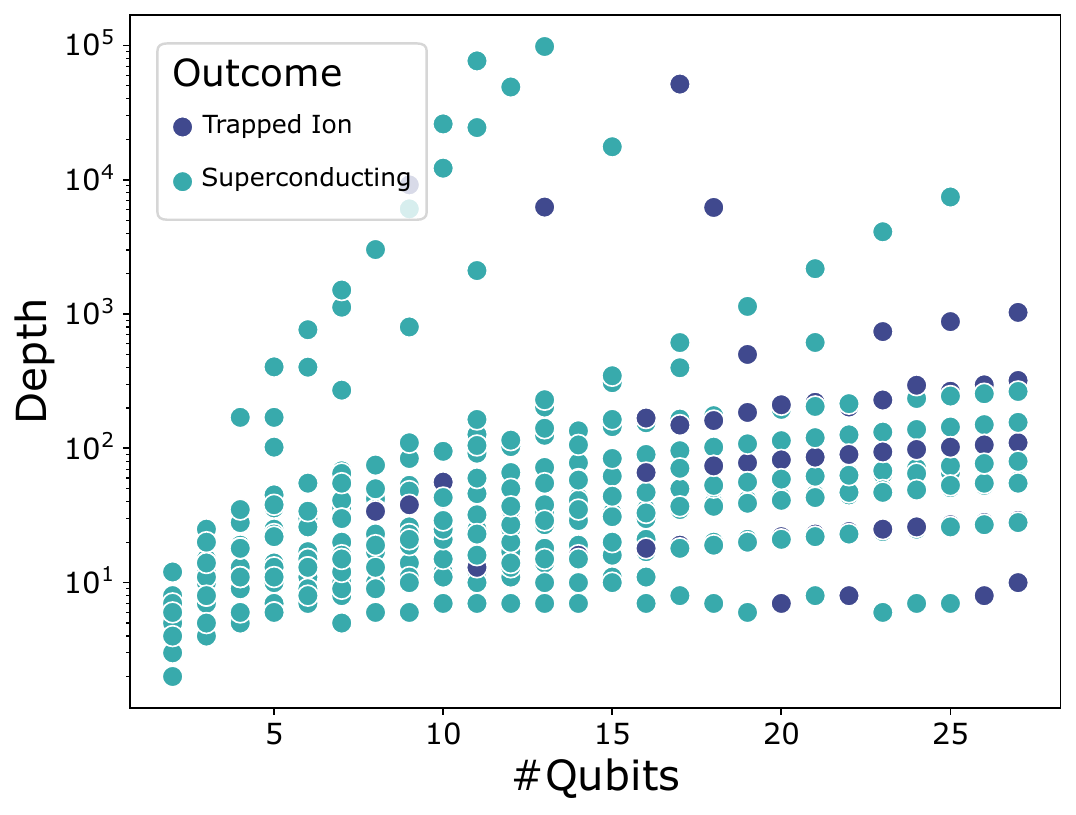}
    \caption{Distribution of circuits based on the number of qubits and circuit depth.}
    \label{fig:qubits_depth}
\end{figure}

Another analysis explores the relationship between circuit depth and the number of gates normalized by the number of qubits (Fig. \ref{fig:depth_nodes_normalized}). Moreover, the correlation between  the number of qubits, the depth was also evaluated.  The results are shown in Fig. \ref{fig:qubits_depth}.
Comparing circuit depth with the number of gates reveals an almost linear relationship between these metrics. However, since the two classes do not clearly separate in the plot, they cannot be distinguished. Similar observations hold when both metrics are normalized by the number of qubits. Lastly, analyzing depth and number of gates as functions of qubit count shows that circuits with more qubits generally exhibit greater depth or complexity (i.e., more operations), but this does not provide any means to distinguish the two classes. 

This analysis proves that considering only the number of qubits, circuit depth, and gate count is insufficient to a-priori determine the optimal device to which implement a quantum circuit. Indeed, the only filtering operation that can be done considering this dataset is to directly assign to the superconducting device the circuit with a number of qubits up to 7. For these reasons, it is not feasible to algorithmically determine the optimal target technology for executing a quantum circuit. Instead, it is necessary to employ ML techniques to automatically identify patterns that can support this decision-making process. Moreover, since the two classes are not linearly separable, one must either extract more informative features or encode the entire quantum circuit into a ML model, allowing it to autonomously learn and extract the most relevant parameters.

\subsection{Models}
For this task, we have focused on the GNNs, a class of neural networks particularly suited for processing graph-structured data. To identify the best-performing model, an analysis was conducted by varying the key hyper-parameters.
The evaluated models consist of a first convolution layer employing either a GAT or GCN layer, followed by one or two residual GCN layers, and conclude with a feed-forward neural network.

The choice to employ a GCN in a ResNet configuration --- which involves adding the output of the convolution to the input data --- aims to mitigate the vanishing gradient problem and enhance model stability.

For models including GAT, the number of hidden dimensions can be set to either $32$ or $64$, while the number of attention heads can be $4$ or $8$.

For models where a GCN layer is used as the first layer, the number of hidden dimensions can be selected from $\{32, 64, 128\}$.

After the graph-based layers, a \textit{Global Mean Pooling} operation is applied to embed the graph into a fixed-size vector. This operation, which represents an hyper-parameter for the model, consists on averaging node features across the node dimension.

The results are then passed into a FFNN composed of $2$ or $3$ hidden layers. The number of units in each layer follows a power-of-two scheme:

\begin{itemize}
    \item The first hidden layer can have a number of units chosen from $\{32, 64, 128, 256, 512, 1024, 2048\}$.
    \item Each subsequent hidden layer can have a number of units that is a power of two, ranging from $16$ up to the size of the preceding layer.
\end{itemize}

Each hidden layer is followed by a \textit{Leaky ReLU} activation function.

The final layer consists of $2$ units with a \textit{softmax} activation function to produce a normalized output vector such that the sum of its elements equals $1$.

For network optimization, due to the class imbalance in the dataset, a balancing technique is applied by adjusting the weights in the loss function according to the number of samples in each class.

The models are trained for $50$ epochs using the cross-entropy loss function, and the optimizer used is \textit{Adam}.

Given the computational cost associated with training GNNs, training is performed on a server equipped with \textbf{4$\times$ NVIDIA Tesla V100 SXM2 GPUs} and \textbf{2$\times$ Intel Xeon Scalable Processors Gold 6130} (2.10 GHz, 16 cores). This setup allows training time for each model to be reduced to just a few seconds.

\section{Results}\label{sec:Results}

To evaluate the performance of our approach, we trained multiple GNN architectures with varying configurations. The dataset was randomly split five times  (5-fold validation) using an 80/20 train-test ratio while maintaining class balance, through a stratified splitting methodology. We report the mean and standard deviation of performance metrics over these splits.

To estimate the model's performance, we considered the overall accuracy and the F1 score for both the minority and majority classes. Table \ref{tab:res-best-gnn} presents the results of the top model, selected based on the highest F1 scores for the minority class.
\renewcommand{\arraystretch}{1.3}
\begin{table*}[t]
    \centering
    \resizebox{\textwidth}{!}{%
    \begin{tabular}{|c|c|c|c|c|c|c|}
        \hline
        \textbf{Model} & \textbf{F1 Class 0 } & \textbf{F1 Class 1 } & \textbf{Acc Test} & \textbf{F1 Class 0  Std }& \textbf{F1 Class 1  Std} & \textbf{Acc Std} \\
        \hline
        GAT\_1GCN\_2FFNN\_32\_4\_2048\_2048\_32 & 0.8556 & 0.9653 & 0.9440 & 0.0152 & 0.0034 & 0.0055 \\
        \hline
    \end{tabular}%
    }\vspace{5pt}
    \caption{Performance metrics of the best GNN model on the test set. \vspace{-10pt}}
    \label{tab:res-best-gnn}
\end{table*}

These results indicate that the best-performing model is the GAT with a hidden dimension of 32, 4 attention heads, 1 GCN in ResNet configuration, and a fully connected network with three hidden layers, where the number of units in each layer is 2048, 2048, and 32, respectively. This model achieved an F1 score of 85.56\% on the minority class and an overall test accuracy of 94.4\%. 

These outcomes can also be compared with those obtained from the framework developed in \cite{quetschlich2024mqtpredictor}, where the best model achieved 98\% top-3 solutions for choosing the best compiler options. However, our analysis differed in that the T-KET compiler was not included and only the 127-superconductor and trapped-ion devices were evaluated. Instead, our focus was on developing a model capable of identifying the optimal device for executing a quantum circuit without any preliminary analysis. Nonetheless, the encouraging results indicate that further refinements could improve its classification performance.

\begin{figure}
    \centering    \includegraphics[width=\linewidth]{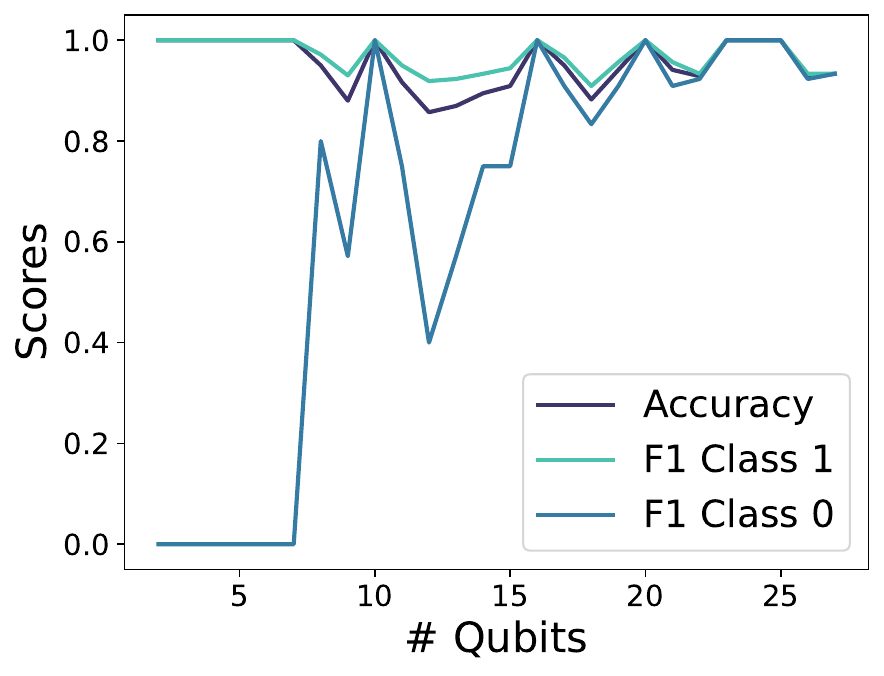}
    \caption{Results of the first best model reported in terms of Accuracy and F1 score for both the classes varying the number of qubits.}
    \label{fig:results-GAT-2048-2048-32}
\end{figure}

Then, we analyze how the models performance as a function of the number of qubits as shown in Figures \ref{fig:results-GAT-2048-2048-32}. 

The results indicate that the model encounter issues for circuits with 8–15 qubits. This is because the number of samples labeled as class 0 is relatively lower compared to cases with more than 15 qubits. This outcome is expected, as Figure \ref{fig:DistibutionVSQubits} illustrates that the Trapped Ion class is significantly underrepresented in this range. Additionally, the high accuracy in the 2–7 qubit interval can be attributed to the absence of any samples from the Trapped Ion class.

\section{Conclusion}\label{sec:Conclusions}

This work introduces a new GNN-based approach for predicting the most suitable quantum hardware platform for a given quantum circuit, formulated as a binary classification task. By exploiting DAG representations of circuits and avoiding manual feature extraction, the proposed model effectively captures structural information relevant to the compilation performance across diverse hardware types.
In our initial analysis, we evaluated 498 circuits, each comprising up to 27 qubits. Each circuit was compiled across all analyzed quantum devices using various compilation options, allowing us to identify the optimal device for execution.
Subsequently, we trained several GNNs by varying the model's architecture and hyperparameters. Our best model achieves 94.4\% accuracy and 85.6\% F1 score on the underrepresented class, proving robust generalization and strong predictive power. These results support the feasibility of integrating graph-based machine learning into quantum software workflows to accelerate and automate compilation-related decisions.

While current results are promising, several roads remain open for improvement. Future efforts will focus on expanding the dataset to include a broader spectrum of hardware types and compiler toolchains, enabling multi-class and multi-label predictions. Additionally, we plan to enhance circuit representations with richer structural and temporal features and explore alternative learning objectives such as ranking or regression for finer-grained predictions.

Finally, we aim to develop an efficient and accurate predictor that assists quantum software engineers and non-expert users in selecting optimal backends for circuit execution, minimizing resource usage while maximizing fidelity and performance in NISQ-era devices. This work represents the first milestone in developing a comprehensive framework to automate the compilation process by exploiting GNNs.

\section*{Acknoledgement}
We would like to thank HPC@POLITO --- a project of Academic Computing within the Department of Control and Computer Engineering at the Politecnico di Torino (\url{http://hpc.polito.it}) --- for providing computational resources.
\bibliographystyle{ieeetr}
\bibliography{sn-bibliography}

\begin{thebibliography}{10}

\bibitem{Preskill2018}
J.~Preskill, ``Quantum computing in the nisq era and beyond,'' {\em Quantum}, vol.~2, p.~79, 2018.
\newblock \url{https://doi.org/10.22331/q-2018-08-06-79}.

\bibitem{Biamonte2017}
J.~B. et~al., ``Quantum machine learning,'' {\em Nature}, vol.~549, pp.~195--202, 2017.
\newblock \url{https://doi.org/10.1038/nature23474}.

\bibitem{Montanaro2016}
A.~Montanaro, ``Quantum algorithms: An overview,'' {\em npj Quantum Information}, vol.~2, p.~15023, 2016.
\newblock \url{https://doi.org/10.1038/npjqi.2015.23}.

\bibitem{Murali2019}
P.~M. et~al., ``Noise-adaptive compiler mappings for noisy intermediate-scale quantum computers,'' in {\em ASPLOS}, pp.~1015--1029, 2019.
\newblock \url{https://doi.org/10.1145/3297858.3304075}.

\bibitem{tannu2019not}
S.~S. Tannu and M.~K. Qureshi, ``Not all qubits are created equal: A case for variability-aware policies for nisq-era quantum computers,'' in {\em Proceedings of the twenty-fourth international conference on architectural support for programming languages and operating systems}, pp.~987--999, 2019.
\newblock \url{ https://doi.org/10.1145/3297858.3304007}.

\bibitem{Wright2019}
K.~W. et~al., ``Benchmarking an 11-qubit quantum computer,'' {\em Nature Communications}, vol.~10, p.~5464, 2019.
\newblock \url{ https://doi.org/10.1038/s41467-019-13534-2}.

\bibitem{Debnath2016}
S.~D. et~al., ``Demonstration of a small programmable quantum computer with atomic qubits,'' {\em Nature}, vol.~536, pp.~63--66, 2016.
\newblock \url{ https://doi.org/10.1038/nature18648}.

\bibitem{quetschlich2024mqtpredictor}
N.~Quetschlich, L.~Burgholzer, and R.~Wille, ``{MQT Predictor: Automatic Device Selection with Device-Specific Circuit Compilation for Quantum Computing},'' {\em ACM Transactions on Quantum Computing}, vol.~6, no.~1, pp.~1--26, 2025.
\newblock \url{https://doi.org/10.1145/3673241}.

\bibitem{quetschlich2023mqt}
N.~Quetschlich, L.~Burgholzer, and R.~Wille, ``Mqt bench: Benchmarking software and design automation tools for quantum computing,'' {\em Quantum}, vol.~7, p.~1062, 2023.
\newblock \url{https://doi.org/10.22331/q-2023-07-20-1062}.

\bibitem{meijer2025comparison}
A.~Meijer-van~de Griend, ``A comparison of quantum compilers using a dag-based or phase polynomial-based intermediate representation,'' {\em Journal of Systems and Software}, vol.~221, p.~112224, 2025.
\newblock \url{https://doi.org/10.1016/j.jss.2024.112224}.

\bibitem{liang2022survey}
F.~Liang, C.~Qian, W.~Yu, D.~Griffith, and N.~Golmie, ``Survey of graph neural networks and applications,'' {\em Wireless Communications and Mobile Computing}, vol.~2022, no.~1, p.~9261537, 2022.
\newblock \url{https://doi.org/10.1155/2022/9261537}.

\bibitem{kipf2017semisupervisedclassificationgraphconvolutional}
T.~N. Kipf and M.~Welling, ``Semi-supervised classification with graph convolutional networks,'' 2017.
\newblock \url{ https://doi.org/10.48550/arXiv.1609.02907 }.

\bibitem{veličković2018graphattentionnetworks}
P.~Velickovic, G.~Cucurull, A.~Casanova, A.~Romero, P.~Lio, Y.~Bengio, {\em et~al.}, ``Graph attention networks,'' 2017.

\bibitem{10.1145/3224206}
E.~M. Elmandouh and A.~G. Wassal, ``Guiding formal verification orchestration using machine learning methods,'' {\em ACM Trans. Des. Autom. Electron. Syst.}, vol.~23, Aug. 2018.
\newblock \url{https://doi.org/10.1145/3224206}.

\bibitem{haaswijk2018deep}
W.~Haaswijk, E.~Collins, B.~Seguin, M.~Soeken, F.~Kaplan, S.~S{\"u}sstrunk, and G.~De~Micheli, ``Deep learning for logic optimization algorithms,'' in {\em 2018 IEEE International Symposium on Circuits and Systems (ISCAS)}, pp.~1--4, IEEE, 2018.
\newblock \url{https://doi.org/10.1109/ISCAS.2018.8351885}.

\bibitem{10213402}
M.~U. Jamal, Z.~Li, M.~T. Lazarescu, and L.~Lavagno, ``A graph neural network model for fast and accurate quality of result estimation for high-level synthesis,'' {\em IEEE Access}, vol.~11, pp.~85785--85798, 2023.
\newblock \url{https://doi.org/10.1109/ACCESS.2023.3303840}.

\bibitem{10.1145/3400302.3415690}
A.~Agnesina, K.~Chang, and S.~K. Lim, ``Vlsi placement parameter optimization using deep reinforcement learning,'' in {\em Proceedings of the 39th International Conference on Computer-Aided Design}, ICCAD '20, (New York, NY, USA), Association for Computing Machinery, 2020.
\newblock \url{https://doi.org/10.1145/3400302.3415690}.

\bibitem{10.1145/3508352.3549346}
K.~Baek, H.~Park, S.~Kim, K.~Choi, and T.~Kim, ``Pin accessibility and routing congestion aware drc hotspot prediction using graph neural network and u-net,'' in {\em Proceedings of the 41st IEEE/ACM International Conference on Computer-Aided Design}, ICCAD '22, (New York, NY, USA), Association for Computing Machinery, 2022.
\newblock \url{https://doi.org/10.1145/3508352.3549346}.

\bibitem{quetschlich2023predicting}
N.~Quetschlich, L.~Burgholzer, and R.~Wille, ``Predicting good quantum circuit compilation options,'' in {\em 2023 IEEE International Conference on Quantum Software (QSW)}, pp.~43--53, IEEE, 2023.
\newblock \url{https://doi.org/10.1109/QSW59989.2023.00015}.

\bibitem{https://doi.org/10.1002/qute.202300128}
A.~Russo, M.~Simoni, D.~Volpe, G.~A. Cirillo, and M.~Graziano, ``Exploring the advantages of layout procedure with fully-connected quantum computing technologies,'' {\em Advanced Quantum Technologies}, vol.~7, no.~1, p.~2300128, 2024.
\newblock \url{https://doi.org/10.1002/qute.202300128}.

\end{thebibliography}

\end{document}